# Technology developments for a scalable heterodyne MMIC array at W-band

Matthew Sieth[1,2], Sarah Church[1,2], Judy M. Lau[1,2], Patricia Voll[1,2], Todd Gaier[3], Pekka Kangaslahti[3], Lorene Samoska[3], Mary Soria[3], Kieran Cleary[4], Rohit Gawande[4], Anthony C.S. Readhead[4], Rodrigo Reeves[4], Andrew Harris[5], Jeffrey Neilson[6], Sami Tantawi[6] and Dan Van Winkle[6]

[1]Department of Physics, Stanford University, 382 Via Pueblo Mall, Stanford, CA 94305, USA
[2]Kavli Institute for Particle Astrophysics & Cosmology, MS 29, Stanford, CA 94309, USA
[3]Jet Propulsion Laboratory, California Institute of Technology, 4800 Oak Grove Drive, Pasadena, CA 91109, USA
[4]Department of Astronomy, California Institute of Technology, 1200 East California Boulevard, Pasadena, CA 91125, USA
[5]Department of Astronomy, University of Maryland, College Park, MD 20742, USA
[6]SLAC National Accelerator Laboratory, 2575 Sand Hill Road, Menlo Park, CA 94025, USA

*Abstract*—We report on the development of W-band (75-110 GHz) heterodyne receiver technology for large-format astronomical arrays. The receiver system is designed to be both mass-producible, so that the designs could be scaled to thousands of receiver elements, and modular. Most of the receiver functionality is integrated into compact Monolithic Microwave Integrated Circuit (MMIC) amplifier-based multichip modules. The MMIC modules include a chain of InP MMIC low-noise amplifiers, coupled-line bandpass filters and sub-harmonic Schottky diode mixers. The receiver signals will be routed to and from the MMIC modules on a multilayer high frequency laminate, which includes splitters, amplifiers, and frequency doublers. A prototype MMIC module has exhibited a band-averaged noise temperature of 41 K from 82-100 GHz and a gain of 29 dB at 15 K, which is the state-of-the-art for heterodyne multi-chip modules.

*Index Terms*—Microwave radiometry, Multichip module, MMICs, Millimeter wave technology

## I. Introduction

Large-format MMIC heterodyne arrays, with hundreds to thousands of receivers, have many exciting applications in millimeter-wave astronomy. Examples of science that would be enabled with these arrays include:

- Probing the process of star formation in galactic molecular gas clouds by mapping gas-tracing spectral lines.
- Mapping low surface brightness emission, like large-scale structure, with existing millimeter-wave interferometers (e.g. CARMA [1]) that could be enhanced with small focal plane arrays on each telescope that would increase the field of view.
- Exploring the history of the early universe via measurements of the polarization of the Cosmic Microwave Background with a short-baseline interferometer [2].
- Examining the expansion rate of the universe by mapping the Sunyaev-Zel'dovich Effect in galaxy clusters with large-format interferometers.

The requirements for all of these applications are similar: low noise, large bandwidth ($\geq$ 15 GHz), and a large number of receiving elements (typically hundreds to thousands).

Previous instruments have already demonstrated the utility of hand-built MMIC heterodyne arrays with tens of elements (e.g. SEQUOIA [3]). The focus of the work presented here is to develop heterodyne receivers whose assembly techniques are amenable to the fabrication of much larger arrays. The use of MMIC (Monolithic Microwave Integrated Circuit) technology allows for all of the receiver components to be integrated into a single compact package, or MMIC module, that is easily mass-producible (see e.g. [4]). The MMIC module interfaces were chosen to maximize modularity so that malfunctioning receivers can be easily swapped. Multilayer boards provide a cost-effective way to carry a high density of signals to and from the MMIC modules.

## II. MMIC Module Designs

Two W-band heterodyne MMIC module designs are presented: a single-sideband (SSB) receiver and a double-sideband (DSB) I-Q receiver. The SSB receiver was designed as a proof-of-concept for a large-scale Cosmic Microwave Background interferometer [2], while the DSB receiver is intended for spectroscopy and builds upon previous work [5].

Both modules include a chain of InP MMIC low-noise amplifiers (LNAs), followed by a sub-harmonic Schottky diode mixer, which downconverts the RF signal to the intermediate frequency (IF) band. The RF input to the modules is WR-10 waveguide, which has a nominal frequency range of 75-110 GHz, but in practice does not become over-moded until 118 GHz. A waveguide-to-microstrip transition [6] precedes the LNAs and planar filters located after the LNAs provide band definition. The component chips are epoxied[1] into 0.381 mm cavities in a base block made from gold-plated brass (see Fig. 1) with a lid (not shown) that forms an RF-tight seal. The components are interconnected via gold ribbon wire bonds. While these MMIC receiver modules were both hand-assembled, their split-block design is conducive to future automated assembly.

The LNAs and mixer are biased by means of hermetic feed-through pins[2] that connect to a PCB with bias circuitry on the reverse side of the block. A nano-miniature 9-pin strip connector provides an easy means of connecting and disconnecting the bias lines.

---
[1]EPO-TEK H20E Silver Epoxy
[2]Thunderline-Z TL1946

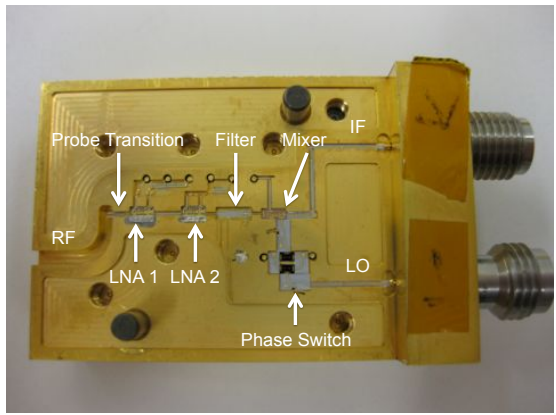

(a) SSB MMIC Module

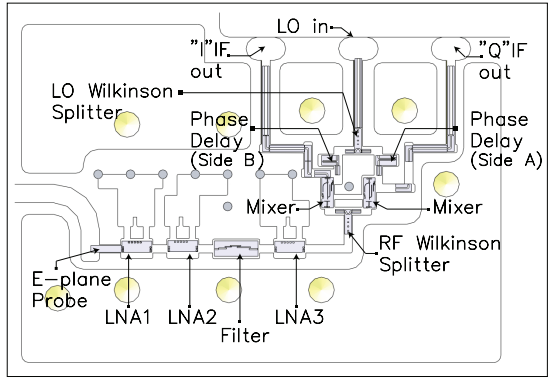

(b) DSB MMIC Module

Fig. 1. (a) Photograph of the single sideband prototype MMIC module with the component chips labeled. Shown are the waveguide-to-microstrip transition, MMIC low-noise amplifiers, Schottky diode mixer, phase switch, and coaxial connectors. (b) Design drawing for the DSB MMIC receiver module. The module includes three low-noise amplifiers, a filter, an I-Q downconversion system and miniature push-on connectors for the IF and LO.

*A. SSB Module*

The SSB prototype receiver module was designed for operation at a fixed LO frequency of 40 GHz with instantaneous RF bandwidth from 85-100 GHz. The second harmonic mixer downconverts the RF signal to a 5-20 GHz intermediate frequency (IF) band. The LO and IF interface is achieved through standard 1.85 mm and 2.92 mm coaxial connectors[3], which were chosen for ease of testing the prototype. The SSB MMIC module also includes an LO phase shifter, which would be required for interferometric operation.

*B. DSB Module*

The DSB MMIC module was implemented using an I-Q configuration, which allows for the downconversion of larger bandwidths at the expense of having to route twice as many IF signals. The I-Q downconversion is implemented with separate second-harmonic mixers for the in-phase and quadrature outputs. The requisite Wilkinson splitters and phase delays were patterned on alumina. The DSB MMIC module is intended to have a tunable LO signal between 42-50 GHz and accepts RF frequencies in the 85-105 GHz range. Miniature push-on connectors (Section IV-C) were used for the IF and LO in order to enhance modularity and compactness.

[3]Anritsu V102F-R & K102F-R

*C. Chip Sets*

*1) Low-Noise Amplifiers:* The sensitivity of the MMIC modules is dominated by the noise contribution of the low-noise amplifiers. The SSB module has two 4-stage 100 nm gate length InP MMIC LNAs [7] that were fabricated by Northrop Grumman Corporation and were also used in the QUIET experiment [4]. The DSB module implements newly developed 35 nm gate length InP MMIC LNAs, which have improved noise properties. Cryogenic noise measurements for this LNA in a waveguide package are shown in Fig. 2 with details of the testing given in Section III. The 35 nm devices have only two stages and 12-15 dB gain per amplifier. A third LNA is required in the DSB module in order to avoid significant noise contributions from the mixer and IF amplifiers. An LNA with higher gain and noise temperature was utilized for this.

*2) Mixers:* Both modules incorporate the same second-harmonic Schottky diode mixer (the DSB module uses two), which has a conversion loss of 12-20 dB. The LO signal for this mixer is sub-harmonically pumped at 40-50 GHz. It is possible to generate and route this signal much more efficiently than the 80-100 GHz signal that would be needed for a fundamental mixer. The mixer is the bandwidth limiting component for both MMIC modules.

*3) Filters:* The receiver modules include 4-pole coupled line bandpass filters, which were implemented on 0.1 mm alumina substrate. In the DSB module, the filter is between the 2nd and 3rd LNAs in order to minimize out-of-band loading at the 3rd LNA and mixer. In the SSB module, the filter is responsible for sideband separation.

## III. MMIC MODULE PERFORMANCE

Noise temperature and gain measurements of the SSB prototype module have been performed using the standard Y-factor technique where a variable thermal load (VTL) was implemented with a waveguide termination and heater. A piece of stainless steel waveguide provides a thermal break between the VTL and MMIC module. The resulting cryogenic data are shown in Fig. 2. The physical temperature was 15 K and the IF was fixed at 1 GHz due to equipment limitations. An average noise temperature of 41 K and gain of 29 dB were obtained across the 82-100 GHz band.

Characterization of the DSB MMIC module was in progress at the time of this writing. This module is expected to exhibit improved noise temperature at cryogenic temperatures based on measurements of the front-end LNA, which is also shown in Fig. 2 and made use of the VTL method described above.

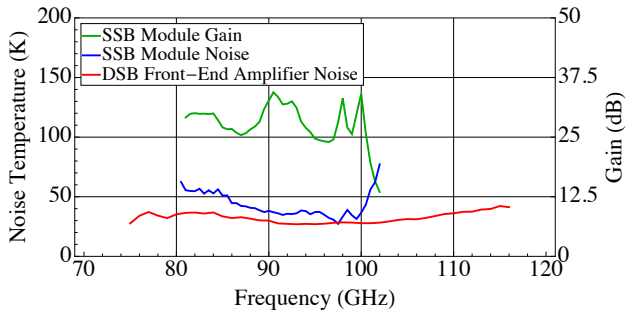

Fig. 2. Noise temperature and gain of the single-sideband prototype MMIC module at 15 K. The measurement was taken with the IF fixed at 1GHz because of equipment limitations. The bandwidth is limited by the mixer performance. Also shown is the cryogenic noise temperature of a newly designed low-noise amplifier, which uses 35 nm gate length InP technology. This amplifier has been installed in the DSB MMIC receiver module.

## IV. Signal Routing

Routing the IF and LO signals is a major challenge for large-format receiver arrays. Coaxial cables and waveguide offer excellent broadband RF performance and isolation between signal paths, but they are bulky and cumbersome to install. Instead, we have developed multilayer PCBs using a high frequency laminate[4]. These boards provide a very cost-effective and compact way to route large numbers of signals to and from the array.

The signal routing and processing will be carried out with two separate boards; a board at approximately 70 K will route the IF and LO signals and will provide IF amplification and LO frequency doubling while a board at around 20 K will route IF and LO signals and will split the LO. Microstrip lines using flex circuit technology will provide the inter-board connections and thermal breaks. A concept drawing illustrating the signal routing is shown in Fig. 3. The multilayer routing boards were prototyped in subsections: a transition from the connectors and surface mount components to intermediate layers, the LO splitter, the miniature push-on connectors, and the flex lines. Each of these prototype subsections are detailed below.

### A. Intermediate Frequency Routing

The IF routing was designed to accommodate not just the DC to 10-15 GHz IF signals from the MMIC receiver modules described in Section II, but also to permit the fabrication of future instruments that can cover 40 GHz instantaneous bandwidth (i.e. two 20 GHz IF signals for an I-Q system).

In order to reduce pixel-to-pixel cross-talk, the IF signals have been routed on one of the interior layers of the board using stripline. The connectors and surface mount components are accessed via stripline-to-microstrip transitons, which are illustrated in Fig. 4. These transitions employ coaxial vias [8], where the signal via is surrounded by a number of ground vias, forming a coaxial transmission line.

[4]Rogers 4350B

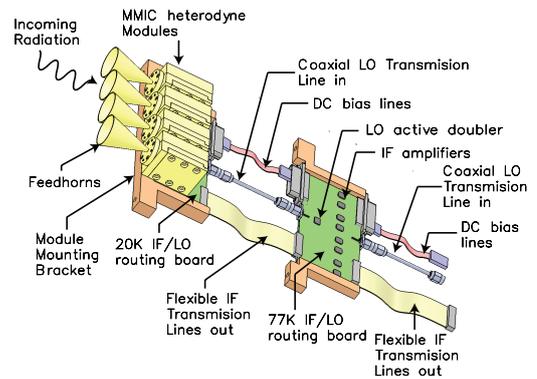

Fig. 3. A concept drawing of the signal routing scheme. Multilayer PCBs route the IF and LO signals to and from the modules. The LO will be routed on the top layer and the IF will be on an interior layer, which improves signal isolation. A 20 K board will include LO splitters and interfaces with the MMIC modules via miniature push-on connectors. A 70 K signal routing board will contain surface mounted IF amplifiers and LO frequency doublers. Thermal breaks are implemented on flex circuitry.

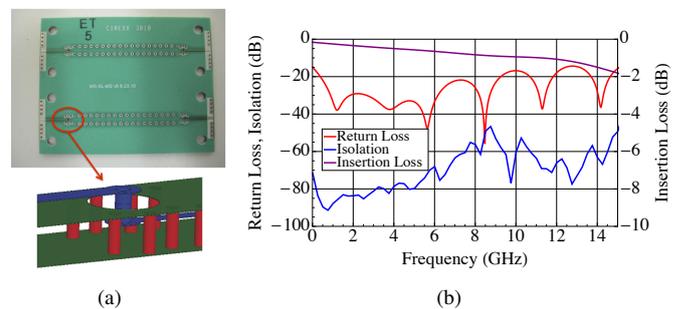

(a)  (b)

Fig. 4. The IF will be routed on an interior layer of the routing boards for improved isolation. A photograph of a stripline-to-microstrip test board is shown in (a). Each line has back-to-back stripline-to-microstrip transitions, where the lines are separated by 15 mm. The total board length is 38.1 mm. The measured performance of stripline-to-microstrip test board is plotted in (b). The worst case return loss is 14.4 dB over the DC-15 GHz band. The insertion loss is less than 2 dB and the isolation is better than 45 dB.

### B. Local Oscillator Routing

The LO signals will also be routed on high frequency laminate on which a 20-25 GHz input signal will be doubled in frequency and then split and routed to four different MMIC modules. A prototype design exists for which the doubling is accomplished via a commercial surface mount component[5]. The splitter will be a rat-race hybrid, which was chosen because it is easier to fabricate than Wilkinson splitters, which require small traces and a very compact resistor. The isolated port of the rat-race splitter will be terminated with a via-less "dot" termination [9] that eliminates the effect of ground vias that have a significant electrical path length at 40-50 GHz.

### C. Miniature Push-On Connectors

The high frequency circuit boards mate to the modules with miniature push-on connectors[6], which are pictured in Fig. 5 along with a plot of the connector performance. These connectors allow for simple connection and disconnection, which would enable malfunctioning receivers in a large-format array to be easily swapped, ensuring that there are no dead pixels. The connectors are also compact, which is conducive to designing small MMIC modules. Note that the mechanical support for the MMIC modules is not through the boards, but will be accomplished through a copper mounting structure.

[5]Mimix XX1000-QT
[6]Corning Gilbert GPPO Interconnect Series

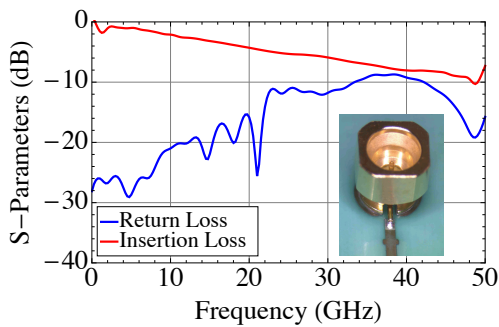

Fig. 5. A plot of the measured performance of a single miniature push-on connector that mates to microstrip. The microstrip section is 76.2 mm, which will be the approximate length of the signal routing boards. The return loss from DC-15 GHz is better than 19 dB and from 42-50 GHz is better than 9.8 dB. The insertion loss has a maximum of 10.3 dB across the band. The inset shows a photograph of the connector with a butterfly stub match.

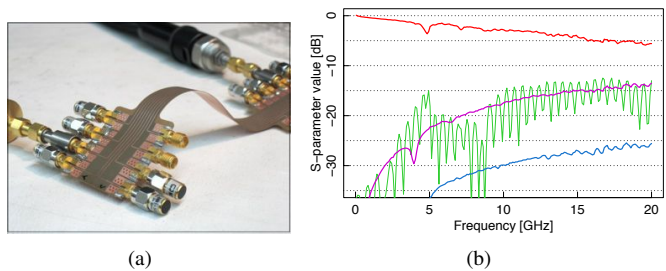

Fig. 6. (a) Photograph of a microstrip test structure with lines using flex circuit technology. The SMA connectors are for testing only. (b) Plot of the measured insertion loss (top line, red), return loss (green), and coupling to closest and next-nearest neighbors (middle, magenta, and bottom, blue curves) of a center line.

*D. Routing Signals Across Thermal Breaks*

The thermal breaks in the IF are implemented with microstrip transmission lines in flex circuit technology. A proof-of-concept test structure with a set of 8 transmission lines has been constructed as shown in Fig. 6. The lines are 0.5 oz copper with 0.028 mm trace widths on 0.127 mm Kapton substrate. The corresponding copper cross section is sufficiently small so that the heat flow through the lines is acceptable. The line spacing is 2.54 mm, which yields better than 15 dB cross talk between adjacent lines across the DC-15 GHz band. Complete S-parameter data for the test structure are shown in Fig. 6. The length of each flex line is expected to be no longer than 30 cm, for which the insertion loss would be approximately 5.5 dB. The flex lines will be wire bonded directly to the routing boards and the cryostat wall connector.

## V. CONCLUSION

Two W-band heterodyne MMIC receiver modules have been developed: a SSB prototype module and a DSB module that will be part of a spectroscopic focal plane array. Across the 82-100 GHz band, the prototype SSB MMIC module has exhibited 41 K noise temperature and 29 dB gain when cooled to 15 K. The DSB module has been fabricated and awaits testing as of this writing. Future designs will focus on extending the bandwidth of the DSB receiver to 75-115.5 GHz.

Signal routing to and from the MMIC modules is implemented with multilayer high frequency PCBs and flex circuitry for the thermal breaks. Routing boards with adequate insertion loss and isolation have been demonstrated, which are both economical and scalable to large-format arrays. The integration of a 4-pixel receiver subarray, including DSB MMIC modules and multilayer routing boards, is currently underway.

## ACKNOWLEDGMENT

The authors would like to thank Matt Morgan for providing the mixer design that was used in the MMIC modules, Heather Owen for invaluable help with assembly, and Richard Lai and Gerry Mei of Northrop Grumman Corporation for their development of InP HEMT processes.

This research was carried out in part by the Jet Propulsion Laboratory (JPL), California Institute of Technology, under a contract with the National Aeronautics and Space Administration. This work was supported by the JPL Strategic University Partnership Program, the National Science Foundation under grant AST-0905855, and Department of Energy contract DE-AC03-76SF00515. P.V. thanks the Harriett G. Jenkins Predoctoral Fellowship Program for their support.

## REFERENCES

[1] (2011) The CARMA website. [Online]. Available: http://www.mmarray.org/
[2] M. Sieth, J. M. Lau, P. Voll, S. Church, P. Kangaslahti, L. Samoska, M. Soria, T. Gaier, D. van Winkle, J. Neilson, S. Tantawi, K. Cleary, and A. C. S. Readhead, "Development of MMIC receivers for cosmic microwave background interferometry," in *Society of Photo-Optical Instrumentation Engineers (SPIE) Conference Series*, vol. 7741, Jul. 2010.
[3] (2011) The SEQUOIA website. [Online]. Available: http://www.astro.umass.edu/f̄crao/instrumentation/sequoia/seq.html
[4] D. Samtleben, "Measuring the Cosmic Microwave Background Radiation (CMBR) polarization with QUIET," *Nuovo Cimento B Serie*, vol. 122, pp. 1353–1358, 2007.
[5] P. Kangaslahti, T. Gaier, J. Cooperrider, L. Samoska, M. Soria, I. O'Dwyer, S. Weinreb, B. Custodero, H. Owen, K. Grainge, J. Lau, S. Church, R. Lai, and X. Mei, "Compact, Miniature MMIC Receiver Modules for an MMIC Array Spectrograph," *NASA Tech. Briefs*, no. NPO-46522, pp. 14–15, Dec. 2009.
[6] Y.-C. Leong and S. Weinreb, "Full band waveguide-to-microstrip probe transitions," in *Microwave Symposium Digest, 1999 IEEE MTT-S International*, vol. 4, 1999, pp. 1435 –1438.
[7] S. Weinreb, R. Lai, N. Erickson, T. Gaier, and J. Wielgus, "W-band InP wideband MMIC LNA with 30 K noise temperature," in *Microwave Symposium Digest, 1999 IEEE MTT-S International*, vol. 1, 1999, pp. 101–104.
[8] E. Pillai, "Coax via-A technique to reduce crosstalk and enhance impedance match at vias in high-frequency multilayer packages verified by FDTD and MoM modeling," *Microwave Theory and Techniques, IEEE Transactions on*, vol. 45, no. 10, pp. 1981 –1985, Oct. 1997.
[9] B. Oldfield, "Connector and termination construction above 50 GHz," *Applied Microwave and Wireless*, vol. 13, no. 4, pp. 56–69, 2001.